\documentclass[prl,aps,twocolumn,showpacs,floatfix,floats,nobalancelastpage]
{revtex4}
\usepackage{graphicx,bm}    

\begin{document}
\title{Spin relaxation dynamics of quasiclassical electrons in ballistic quantum dots with strong
spin-orbit coupling}
\author{Cheng-Hung Chang$^1$, A.G. Mal'shukov$^2$, and K.A. Chao$^3$ }
\date{\today}
\affiliation{
$^1$ National Center for Theoretical Sciences,
Physics Division,
Hsinchu 300, Taiwan \\
$^2$ Institute of Spectroscopy, Russian Academy of Science,
142190 Troitsk, Moscow oblast, Russia \\
$^3$
Solid State Theory, Department of Physics, Lund University,
S-223 62 Lund, Sweden}



\begin{abstract}
We performed path integral simulations of spin evolution
controlled by the Rashba spin-orbit interaction in the
semiclassical regime for chaotic and regular quantum dots. The
spin polarization dynamics have been found to be strikingly
different from the D'yakonov-Perel' (DP) spin relaxation in bulk
systems. Also an important distinction have been found between
long time spin evolutions in classically chaotic and regular
systems. In the former case the spin polarization relaxes to zero
within relaxation time much larger than the DP relaxation, while
in the latter case it evolves to a time independent residual
value. The quantum mechanical analysis of the spin evolution
based on the exact solution of the Schr\"{o}dinger equation with
Rashba SOI has confirmed the results of the classical simulations
for the circular dot, which is expected to be valid in general
regular systems. In contrast, the spin relaxation down to zero in
chaotic dots contradicts to what have to be expected from quantum
mechanics. This signals on importance at long time of the
mesoscopic echo effect missed in the semiclassical simulations.

\end{abstract}



\maketitle

\section{I Introduction}

Spin relaxation in semiconductors is an important physical
phenomenon being actively studied recently in connection with
various spintronics applications \cite{reviewonsptronics}. In
doped bulk samples and quantum wells (QW) of III-V semiconductors
at low temperatures spin relaxation is mostly due to the DP
mechanism \cite{DP}. This mechanism does not involve any
inelastic processes, so that the exponential decay of the spin
polarization is determined entirely by the spin-orbit interaction
(SOI) and elastic scattering of electrons on the impurities.
However, in case of confined systems such as quantum dots (QD)
with atomic-like eigenstates, the SOI has been incorporated into
the structure of the wave functions of the discrete energy
levels. Without inelastic interactions, an initial wave packet
with a given spin polarization will evolve in time as a coherent
superposition of these discrete eigenstates. Therefore, the
corresponding expectation value of the spin polarization will
oscillate in time without any decay. To obtain a polarization
decay in the QD's, extra effects have to be introduced into the
system, e.g., the \emph{inelastic} interactions between electrons
and phonons mediated by the spin-orbit \cite{Khaetskii, Woods}
and nuclear hyperfine effects \cite{Khaetskii, Merkulov,
Semenov}. Accordingly, a spin relaxation in QD's induced by these
effects is a real dephasing process.

Unlike such an inelastic relaxation in QD's, the DP spin
relaxation in unbounded systems seems to be a quite different
phenomenon, because the scattering on impurities is elastic and
there is no dephasing of the electron wave functions in the
systems. However, the spin polarization does decay in time
exponentially, as if it would be a true dephasing process. To
explain this phenomenon, let us consider an electron moving
diffusively through an unbounded system with random elastic
scatters. This electron is described by a wave packet represented
by a superposition of \emph{continuum} eigenstates. During a DP
relaxation process, the spin expectation value expressed as a
bilinear combination of these wave amplitudes will decay
exponentially in time. This process can be easily understood from
the semiclassical Boltzmann or Fokker-Plank approach \cite{DP}.
Indeed, keeping in mind that the SOI has the form
$\bm{\sigma}\cdot\mathbf{h(k)}$, where $\bm{\sigma}$ is the
vector, whose components are the three Pauli matrices, and
$\mathbf{h(k)}$ is the effective magnetic field, whose magnitude
and direction depend on the electron momentum $\mathbf{k}$, one
can envision spin relaxation as the spin random walk on the
surface of the unit sphere, similar to that in Fig. 1(c).
Starting at the north pole, the spin precesses around
$\mathbf{h(k_1)}$ until the momentum direction is changed by a
scattering on an impurity. Thereafter, the magnetic field changes
its direction to $\mathbf{h(k_2)}$ and the spin continues its
precession around this new direction. If the spin rotation angle
between successive scattering events is small, the sequence of
such rotations results in a diffusive spreading of the initial
polarization.

Returning to QD's, a natural question emerges: what sort of spin
evolution can be generated by the DP mechanism in a ballistic QD
whose size is much larger than the electron wavelength at the
Fermi surface and where the mean spacing between energy levels is
much less than $\hbar/T$, where $T$ is the mean time between
electron collisions with the boundary? Similar to the example in
Fig. 1, the spin evolution in this semiclassical regime can be
studied by tracking the spin walk on the sphere, when particles
move along the classical trajectories inside the QD's.
Intuitively, one would expect the spin evolution in this case to
be similar to the spin random walk governed by the impurity
scattering in unbounded samples. However, this expected analogy
with the open system is wrong. Indeed, in an unbounded system,
the steps of the random walk are uncorrelated. This results in a
diffusive decay of the spin polarization down to zero for any
nonzero SOI. But in case of QD's, the steps of the random walk on
the sphere are correlated due to the confinement of electron
trajectories within the dots. As we will show below, such
correlations not only lead to a spin relaxation much longer than
the DP relaxation in unbounded systems, but also to a non-zero
final polarization value at long time for certain quantum dot
geometries. Here, we do not take into account the inelastic
mechanisms \cite{Khaetskii, Merkulov, Semenov, Woods} which
always drive the spin polarization to zero in long time. These
mechanisms are assumed to be absent, because they become
inefficient at low sufficiently temperatures.

In this article we carry out a semiclassical analysis of the DP
relaxation in 2-dimensional (2D) QD's of various geometries,
including a circular dot, a triangular dot, a generalized Sinai
billiard, and a  circular dot with diffusive scattering on the
boundary. We focus on the case of the strong SOI, such that the
characteristic spin orbit length $L_{so} \equiv v_F /\hbar
h(\bm{k_F})$ is not much larger than the dot size $L$. Such a
regime can be realized in the InAs based heterostructures for $L
\sim 0.5 -1 \mu m$ \cite{Nitta}. We found that in the short time
scale $\sim T$ the spin relaxation dynamics in all geometries
shares a common feature: After a fast initial drop during the
time interval $\sim T$, the spin polarization continues to
oscillate weakly around some value. For weak SOI with $L_{so}\gg
L$, all residual values for different dot geometries are quite
close to one up to the cutoff time of our numerical simulations
($\sim 10^3 T$). For stronger SOI with $L_{so}\geq L$, the
initial drop of the spin polarization is considerably larger
compared to the weak SOI regime. The spin evolution after that
drop depends on the dot geometry. In the case of circular and
triangular dots, which are examples of systems with regular
classical dynamics, the corresponding spin polarizations approach
nonzero residual values. However, in the case of chaotic and
random systems (e.g., Sinai billiard and circular dot with rough
boundaries, respectively), the spin polarizations slowly decrease
to zero after that initial drop. But this decreasing is much
longer than the DP relaxation in an unbounded system, in which
the mean impurity scattering time is $\sim T$. For very strong
SOI with $L_{so} < L$, the spin polarization after the initial
drop reaches zero and later on oscillates with a large amplitude.

These results clearly demonstrate that the spin evolution in QD's
is qualitatively distinct from the DP spin relaxation in
unbounded systems. In order to elucidate the physical origin of
this phenomenon, two investigations have been performed. First,
the spin evolution along a single electron trajectory was studied
in detail, which provided a clue for understanding the
above-mentioned polarization behavior. Second, the residual
polarization obtained from the classical simulations for a
circular quantum dot was compared with that derived from the
exact solution of the Schr\"{o}dinger equation. A good agreement
between the results from these two approaches has been found.
However, for QD's with chaotic and random electron dynamics, the
general quantum mechanical analysis revealed a contradiction to
the long time spin evolution observed in our semiclassical
simulations.

The article is organized in the following way. In Section II the
general expression of the polarization will be derived for the
spin evolution via classical path integrals. In Section III the
results of the numerical simulations in different quantum dots
will be demonstrated. The quantum mechanical theory for the spin
polarization in the circular quantum dot will be presented in
Section IV, with the calculation in detail shown in the Appendix.
Discussion and conclusion will be given in Section V.

\section{II Path integrals for the spin evolution}

The Hamiltonian of the system,
\begin{equation}\label{eq1}
H=H_0+\bm{\sigma}\cdot\mathbf{h(\hat{k})},
\end{equation}
consists of the spin independent part $H_0$, which is the
electron kinetic energy plus the 2D confining potential
$V(\mathbf{r})$, and the spin-orbit interaction. In III-V
semiconductor heterostructures the effective "magnetic" field
$\mathbf{h(\hat{k})}$ is given by the sum of the Rashba
\cite{Rashba} and the Dresselhaus \cite{Dress} terms. If the
$z$-axis is chosen perpendicular to the heterostructure
interface, the magnetic field $\bm{h}_R$ contributing to the
Rashba term has two components
$(h_{R}^x(\bm{\hat{k}}),h_{R}^y(\bm{\hat{k}}))=(\alpha_R
\hat{k}_y,-\alpha_R \hat{k}_x)$, where $\hbar \bm{\hat{k}}=(\hbar
\hat{k}_x,\hbar \hat{k}_y)$ is the momentum operator. In the 2D
confinement, the magnetic field $\bm{h}_D$ contributing to the
Dresselhaus term contains both linear and cubic parts with
respect to $\bm{\hat{k}}$ \cite{Eppenga}. In a [001] oriented QW
the linear term has the components
$(h_{D}^x(\bm{\hat{k}}),h_{D}^y(\bm{\hat{k}}))=(\alpha_D
\hat{k}_x,-\alpha_D \hat{k}_y)$. For heterostructures with a
typical $\sim$10nm confinement in $z$-direction, the linear part
of $\mathbf{h}_D$ is usually larger than the cubic part, except
the case of high doping concentration \cite{Jusserand}. The
Rashba term is not zero only in heterostructures with asymmetry
in their growth direction. This term can be much larger than the
Dresselhaus term in the narrow gap InAs based systems
\cite{Nitta}. In this article we will study the spin evolution
induced by the Rashba term. But, since the SOI Hamiltonians
corresponding to the Rashba and the linear Dresselhaus terms can
be transformed from one to the other by the unitary matrix
$(\sigma_{x}+\sigma_{y})/\sqrt 2$, our results are also valid for
systems in which the linear Dresselhaus term dominates the SOI.

Let us suppose $E_n$ to be the n-th quantized energy level with
the eigenfunction $\varphi_{n}$, which is a two component spinor.
At zero magnetic field this quantum state is at least doubly
degenerate. Let
\begin{equation}\label{eq2}
\psi(\bm{r})=e^{i\mathbf{k\cdot
r}}\Phi(\mathbf{r}-\mathbf{R})\chi
\end{equation}
be the wave packet created at time $t=0$, centered at the point
$\mathbf{R}$, and propagating with the 2D wavevector
$\mathbf{k}$. The function $\Phi(\mathbf{r})$ is assumed to be
slowly varying within the scale of the electron wavelength
$2\pi/k$ and normalized, so that the integral
$\int|\Phi(\mathbf{r})|^2\,d^2r$ over the QD volume is equal to
$1$. The initial spin polarization
$\mathbf{P}(0)=\sum_{\alpha\beta}\chi_\alpha^*
\bm{\sigma}_{\alpha\beta}\chi_\beta$ is the sum over the two
components $\chi_\alpha$ of the spinor $\chi$, where
$\alpha\in\{1,2\}$. For $t>0$, the wave packet evolves in time as
\begin{equation}\label{eq3}
\psi(\mathbf{r},t) =\sum_n c_n\varphi_n (
\mathbf{r})\,e^{-i\,E_nt/\hbar}\,,
\end{equation}
where
\begin{equation}\label{eq4}
c_n=\int \varphi_n^\dag(\mathbf{r})\psi(\mathbf{r})\,d^2 r\,.
\end{equation}
In terms of $\psi(\mathbf{r},t)$ the  time dependent spin
polarization is expressed as
\begin{equation}\label{eq5}
\mathbf{P}(t)=\sum_{\alpha\beta}\int\psi_{\alpha}^*(\mathbf{r},t)
\bm{\sigma}_{\alpha\beta}\psi_{\beta}(\mathbf{r},t)\, d^2 r\,,
\end{equation}
with three components $\mathbf{P}(t)=(P^x(t),P^y(t),P^z(t))$.

For further analysis it is convenient to introduce the retarded
and advanced Green functions,
\begin{eqnarray}\label{eq6}
&&G_{\alpha\beta}^r(t-t',\mathbf{r,r'})=G_{\beta\alpha}^{a*}(t'-t,\mathbf{r',r}) \\
&&=-i\sum_{n}\varphi_{n\alpha}(\mathbf{r})\varphi_{n\beta}^*(\mathbf{r}')e^{-iE_n
(t-t')}\Theta(t-t')\,, \nonumber
\end{eqnarray}
which are 2$\times$2 matrices acting on the
SU(2) spin space, where $\Theta(t-t')$ is the Heaviside function.
Using these Green functions, the spin-spin correlation function
can be defined as
\begin{eqnarray}\label{eq7}
&& K^{ij}(\mathbf{r,r'};t-t') \\
&&=\int \mbox{Tr}\left[\sigma^i G^r(t-t',\mathbf{r'',r})\sigma^j
G^a(t'-t,\mathbf{r',r''})\right]d^2r''\,, \nonumber
\end{eqnarray}
where $i,j\in\{x,y,z\}$. This definition together with Eqs.
(\ref{eq3}-\ref{eq5}) lead to the expression for the polarization
evolution in time,
\begin{eqnarray}\label{eq8}
P^{i}(t)=\frac{1}{2}\int
K^{ij}(\mathbf{r,r'};t)\Phi(\mathbf{r}-\mathbf{R})
\Phi^{*}(\mathbf{r'}-\mathbf{R}) \nonumber \\
\times e^{i\mathbf{k}(\mathbf{r}-\mathbf{r}')}P^j(0)d^2r d^2r'\,.
\end{eqnarray}

For classical simulations below, the semiclassical approximation
of Eq. (\ref{eq8}) is required. It can be derived from a standard
path integral formalism \cite{Schulman}, by representing the
retarded Green function in Eq. (\ref{eq7}) as the sum of
products,
\begin{eqnarray}\label{eq9}
&&G_{\alpha\beta}^r(t-t',\mathbf{r,r'}) \\
&&=\int d\mathbf{r}_1\cdots d\mathbf{r}_n
\sum_{\alpha_1,\,\alpha_2,\cdots}
\langle\mathbf{r},\alpha|e^{-iH(t-t_1)}|\mathbf{r}_1,\alpha_1\rangle \nonumber\\
&&\langle\mathbf{r}_1,\alpha_1|e^{-iH(t_1-t_2)}|\mathbf{r}_2,\alpha_2\rangle\cdots
\langle\mathbf{r}_n,\alpha_n|e^{-iH(t_n-t')}
|\mathbf{r'},\beta\rangle \nonumber
\end{eqnarray}
of the evolution operators $e^{-iH(t_i-t_j)}$ within the
infinitesimally short time intervals $(t_i-t_j)$. Thereafter, the
Green function can be expressed as the path integral of
$\mathbf{T}\exp\left[\frac{i}{\hbar}S(t-t',\mathbf{r,r'})\right]$,
where the action
\begin{eqnarray}\label{eq10}
&&S(t-t',\mathbf{r,r'}) \\
&&=\int^{t}_{t'}\left[\frac{m^*}{2}v^2(\tau)-V\left(\mathbf{r}(\tau)\right)
-\mathbf{h}_R\left(\frac{m^*\mathbf{v}(\tau)}{\hbar}\right)\cdot\bm{\sigma}\right]d\tau
\,, \nonumber
\end{eqnarray}
is a time integral of the particle Lagrangian evaluated along a
trajectory starting from $\mathbf{r'}$ at time $t'$ and ending
with $\mathbf{r}$ at time $t$, where
$\mathbf{v}(\tau)=\frac{d\mathbf{r}}{d\tau}$. In this Lagrangian,
the constant term $m^* \alpha_R^2 /2$ is ignored, because it only
gives a phase factor. Since the SOI Lagrangians on different parts of
the trajectory do not commute, one has to keep different
$\exp\left[\frac{i}{\hbar}S(t-t',\mathbf{r,r'})\right]$
in the order of the sequence in Eq. (\ref{eq9}), which is preserved by
the time ordering operator $\mathbf{T}$.

By using the saddle point approximation, the path integral in Eq.
(\ref{eq9}) can be reduced to a sum over all classical trajectories
$\gamma$ \cite{Schulman},
\begin{eqnarray}\label{eq11}
&& G^r(t-t',\mathbf{r,r'}) \\
&&=\frac{1}{2\pi}\sum_\gamma\sqrt{J(\mathbf{r,r'})}
e^{\frac{i}{\hbar}S_0(t-t',\mathbf{r,r'})}U(t-t',\mathbf{r,r'})\,,
\nonumber
\end{eqnarray}
with the spin independent monodromy matrix
$J(\mathbf{r,r'})=\det\left(\frac{\partial^2 S_0}{\partial
r_i\partial r'_j}\right)$ and the spin independent classical
action $S_0(t-t',\mathbf{r,r'})$ along the classical trajectories.
The spin dependence part of the Green function is represented
by the unitary matrix
\begin{equation}\label{eq12}
 U(t-t',\mathbf{r,r'})=\mathbf{T}e^{-\frac{i}{\hbar}\int^{t}_{t'}
 \mathbf{h}_R\big(\frac{m^*\mathbf{v}(\tau)}
 {\hbar}\big)\cdot\bm{\sigma}}d\tau\,.
\end{equation}
Such a decoupling of the spatial and spin degrees of freedom can
be done under the assumption that the classical paths are only
weakly perturbed by SOI, which is reasonable, when the SOI
parameter $\alpha_R$ is much less than the electron Fermi
velocity. Under this assumption, all quantities $J$, $S$, and $U$
are evaluated on the unperturbed trajectories.

Inserting Eq. (\ref{eq11}) and (\ref{eq12}) into Eq. (\ref{eq7})
and (\ref{eq8}), we obtain a semiclassical expression for the spin
polarization. This expression can be substantially simplified
after integrating over coordinates $\mathbf{r}$ and $\mathbf{r'}$
in Eq. (\ref{eq8}). Indeed, let us consider the integral in Eq.
(\ref{eq8}),
\begin{eqnarray}\label{eq13}
\int  \sqrt{J(\mathbf{r'',r})}
e^{\frac{i}{\hbar}S_0(t,\mathbf{r'',r})}\Phi
(\mathbf{r}-\mathbf{R}) e^{i\mathbf{k}\mathbf{r}}U(t,\mathbf{r'',r})d^2r.\;\;
\end{eqnarray}
In the semiclassical limit, the exponential function
$\exp\left[\frac{i}{\hbar}S_0(t,\mathbf{r'',r})\right]$ rapidly
oscillates as a function of $\mathbf{r}$ with a period given by
the Fermi wavelength. However, $J$, $U$, and $\Phi$ are slowly
varying functions of $\mathbf{r}$. The length scale of $J$'s
variation is given by the dot size. The spatial changes of $U$
are controlled by the spin orbit length
$L_{so}=\hbar/(m^*\alpha_{R})$, which is assumed to be much
larger than the Fermi wavelength. Therefore, the influence of the
SOI on the saddle-point position can be ignored. The variation of
$\Phi$ also can be ignored, because this function was assumed to
change weakly within the length scale equal to the electron
wavelength. Under these approximations, we obtain the
saddle-point equation in the form
\begin{equation}\label{eq14}
 \frac{\partial S_0(t,\mathbf{r'',r})}{\partial
 \mathbf{r}}+\hbar \mathbf{k}=0\,.
\end{equation}
This equation is the classical equation of motion. It determines
the trajectory
$\mathbf{r}=\mathbf{r}_0(\mathbf{r''}(t),\mathbf{p}(0))$ which
passes through the given point $\mathbf{r}''(t)$ at the instant
$t$, on condition that at $t=0$ the initial momentum was
$\mathbf{p}(0)=\hbar \mathbf{k}$. Therefore, the saddle point
$\mathbf{r}$ is a particle coordinate at $t=0$ belonging to this
trajectory. Since the integral over $\mathbf{r}'$ in Eq.
(\ref{eq8}) is taken around this extremum, the value
$\mathbf{r'}=\mathbf{r}=\mathbf{r}_0$ are inserted into all
slowly varying functions $J$, $U$ and $\Phi$.

Further, to calculate the integral
over $\mathbf{r}$ in Eq. (\ref{eq13}), the action
$S_0(t,\mathbf{r'',r})$ is expanded around
$\mathbf{r}=\mathbf{r}_0$ up to the second order,
\begin{eqnarray}\label{eq15}
&&S_0(t,\mathbf{r'',r})+\hbar
\mathbf{k}=S_0(t,\mathbf{r''},\mathbf{r}_0)\nonumber \\
&&+\frac{1}{2}\frac{\partial
S_0(t,\mathbf{r''},\mathbf{r}_0)}{\partial r^{i}_0\partial
r^{j}_0}(r-r^{i}_0)(r-r^{j}_0)\,.
\end{eqnarray}
The integration over $\mathbf{r}$ and $\mathbf{r'}$ in Eq.
(\ref{eq8}) gives  $(2\pi)^2/\det\left(\frac{\partial
S_0(t,\mathbf{r''},\mathbf{r}_0)}{\partial r^{i}_0\partial
r^{j}_0}\right)$. Combining this Jacobian  with
$J(\mathbf{r'',r}_0)$ we obtain
\begin{eqnarray}\label{eq16}
&&\det\left(\frac{\partial
S_0(t,\mathbf{r''},\mathbf{r}_0)}{\partial r''^{i}\partial
r^{j}_0}\right)\left[\det\left(\frac{\partial
S_0(t,\mathbf{r''},\mathbf{r}_0)}{\partial r^{i}_0\partial
r^{j}_0}\right)\right]^{-1}\nonumber \\
&&=\det\left(\frac{\partial r^{i}_0}{\partial r''^{j}} \right)\,.
\end{eqnarray}
By using the identity
\begin{equation}\label{eq17}
\det\left(\frac{\partial r^{i}_0}{\partial r''^{j}}
\right)d^2r''=d^2r_0\,,
\end{equation}
Eq. (\ref{eq7}) can be integrated over $\mathbf{r}_0$, instead of
$\mathbf{r}''$, which leads to the expression of the semiclassical
spin polarization,
\begin{equation}\label{eq18}
P^i_c(t)=\frac{P^j(0)}{2}\int
R^{ij}(\mathbf{r,r'},t)|\Phi(\mathbf{r'}-\mathbf{R})|^2 d^2r' \,,
\end{equation}
with
\begin{equation}\label{eq19}
R^{ij}(\mathbf{r,r'},t)=\mbox{Tr}\left[\sigma^i
U(t,\mathbf{r,r'})\sigma^j U^{\dag}(t,\mathbf{r,r'})\right]\,.
\end{equation}
Equation (\ref{eq18}) describes the spin evolution of a particle
initially distributed around the point $\mathbf{R}$ with the
probability density $|\Phi(\mathbf{r'}-\mathbf{R})|^2$. This
particle starts its classical motion from the point $\mathbf{r}'$
with the momentum $\hbar \mathbf{k}$ at time zero and arrives in
the position $\mathbf{r}$ at time $t$. In the following, we are
interested in the spin evolution averaged over an ensemble of
electrons with uniformly distributed coordinates $\mathbf{R}$ and
random directions of the initial momenta on the Fermi surface.
After averaging Eq. (\ref{eq18}) over $\mathbf{R}$ and the
angular coordinate $\theta_\mathbf{k}$ of the momentum
$\mathbf{k}$, we obtain the simple expression:
\begin{equation}\label{eq20}
P^i_c(t)=\frac{P^j(0)}{4\pi}\int R^{ij}(\mathbf{r,r'},t) d^2r'
d\theta_\mathbf{k}\,.
\end{equation}
It should be noted that after the integration over $\mathbf{R}$
this expression does not depend on the initial wave packet
envelope $\Phi (\mathbf{r}-\mathbf{R})$. Therefore, the same Eq.
(\ref{eq20}) holds for $\Phi =const$, so that the initial state
can be simply a plane wave.

\section{III Numerical Results}

Equation (\ref{eq20}) is the basic equation for our numerical
simulations of the spin polarization. Below we will restrict
ourselves to the case when the initial polarization
$\mathbf{P}(0)$ is directed along the $z$ axis, so that
$P^z(0)=1$, and the polarization to be calculated at time $t$ is
also in $z$-direction.

\subsection{III.1 Spin evolution in ballistic quantum dots}

Consider a free electron confined inside a quantum dot and moving
along the trajectory $\gamma$, which consists of the successive
straight segments $\gamma_j$ of the lengths $l_j$ with $j=1$, $2$,
$...,\,n$. The spin state along this trajectory can be described
by the evolution operator $U_\gamma=U(t,\mathbf{r,r'})$ in Eq.
(\ref{eq12}) with $t'=0$. This operator can be represented as a
product
\begin{equation}\label{eq21}
U_\gamma=U_{\gamma_n} \cdots U_{\gamma_j} \cdots
U_{\gamma_2}U_{\gamma_1},
\end{equation}
of the individual operators
\begin{equation}\label{eq22}
U_{\gamma_j} = \exp\left[-i\psi_j\,J_j\right],
\end{equation}
with $\psi_j=\frac{l_j}{L_{so}}$, $J_j=\mathbf{N}_j\cdot
\bm{\sigma}$. Thereby, $\mathbf{N}_j={\bm{n}}_j\times {\bm{e}}_z$
is the unit vector parallel to the effective magnetic field
$\mathbf{h(k)}=\alpha_R(\mathbf{k}\times{\bm{e}}_z)$, where
${\bm{n}}_j=\mathbf{k}/|\mathbf{k}|$ is the unit vector along the
trajectory segment $j$ and ${\bm{e}}_z$ is the unit vector in
$z$-direction. Since $J_j$ is a vector in the space of the Pauli
matrices, the individual operator in Eq. (\ref{eq22}) has a
simple form
\begin{equation}\label{eq23}
U_{\gamma_j}=\cos(\psi_j)\,{\bf 1} - i\sin(\psi_j) \, J_j,
\end{equation}
with the identity matrix ${\bf 1}$.

Let us assume the $j$-th segment $\gamma_j$ to have the angle
$w_j$ with respect to the x-axis. Accordingly, the vector
$\mathbf{N}_j$ has the angle $w_j-\pi/2$, so that we get the
explicit expression $J_j=\sin(w_j)\sigma_x-\cos(w_j)\sigma_y$. In
SU(2) representation, the operator $U_{\gamma_j}$ can be
expressed as the matrix
\begin{equation}\label{eq24}
U_{\gamma_j}=\left(\begin{array}{cc} \cos(\psi_j) &
\sin(\psi_j)\,e^{-i\;w_j}
\\ -\sin(\psi_j)\,e^{i\;w_j} & \cos(\psi_j)\end{array}\right),
\end{equation}
which acts on the spin state
\begin{equation}\label{eq25}
\chi=\left(\begin{array}{c} \chi_1 \\ \chi_2 \end{array}\right)
=\left(\begin{array}{c}
\cos(\theta/2)\,e^{i\,\phi_1} \\
\sin(\theta/2)\,e^{i\,\phi_2}
\end{array}\right).
\end{equation}
In SO(3) representation, the operator $U_{\gamma_j}$ corresponds
to a spin rotation around the axis $\mathbf{N}_j$ through the
angle $2\psi_j$. The three components of the spin expectation
value are related to the spinor $\chi$ by
\begin{eqnarray}\label{eq26}
\mathbf{s}=\left(\begin{array}{c} s_x \\ s_y \\ s_z
\end{array}\right)
=\left(\begin{array}{c}
2\, \mbox{Re}(\chi_1^*\chi_2) \\
2\, \mbox{Im}(\chi_1^*\chi_2) \\
|\chi_1|^2-|\chi_2|^2
\end{array}\right).
\end{eqnarray}
For convenience, we will call the vector projections
$s_i\in[-1,1]$ as \emph{spin} components, although they are twice
larger than the corresponding values for the spin 1/2.

\begin{figure}[htbp!]
\begin{minipage}[b]{8.4cm}
\includegraphics[width=8.4cm]{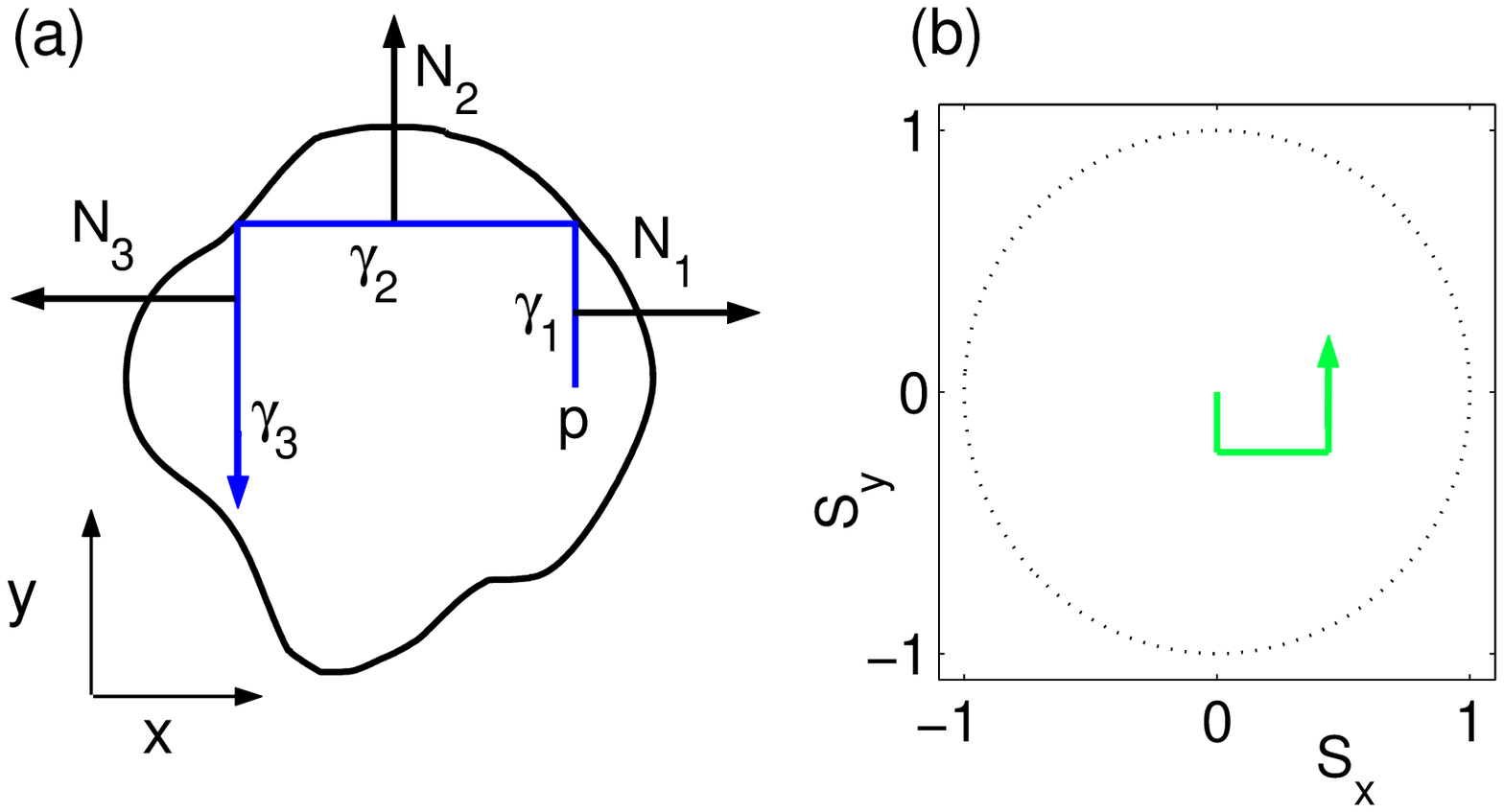}
\end{minipage}
\begin{minipage}[b]{8.4cm}
\includegraphics[width=8.4cm]{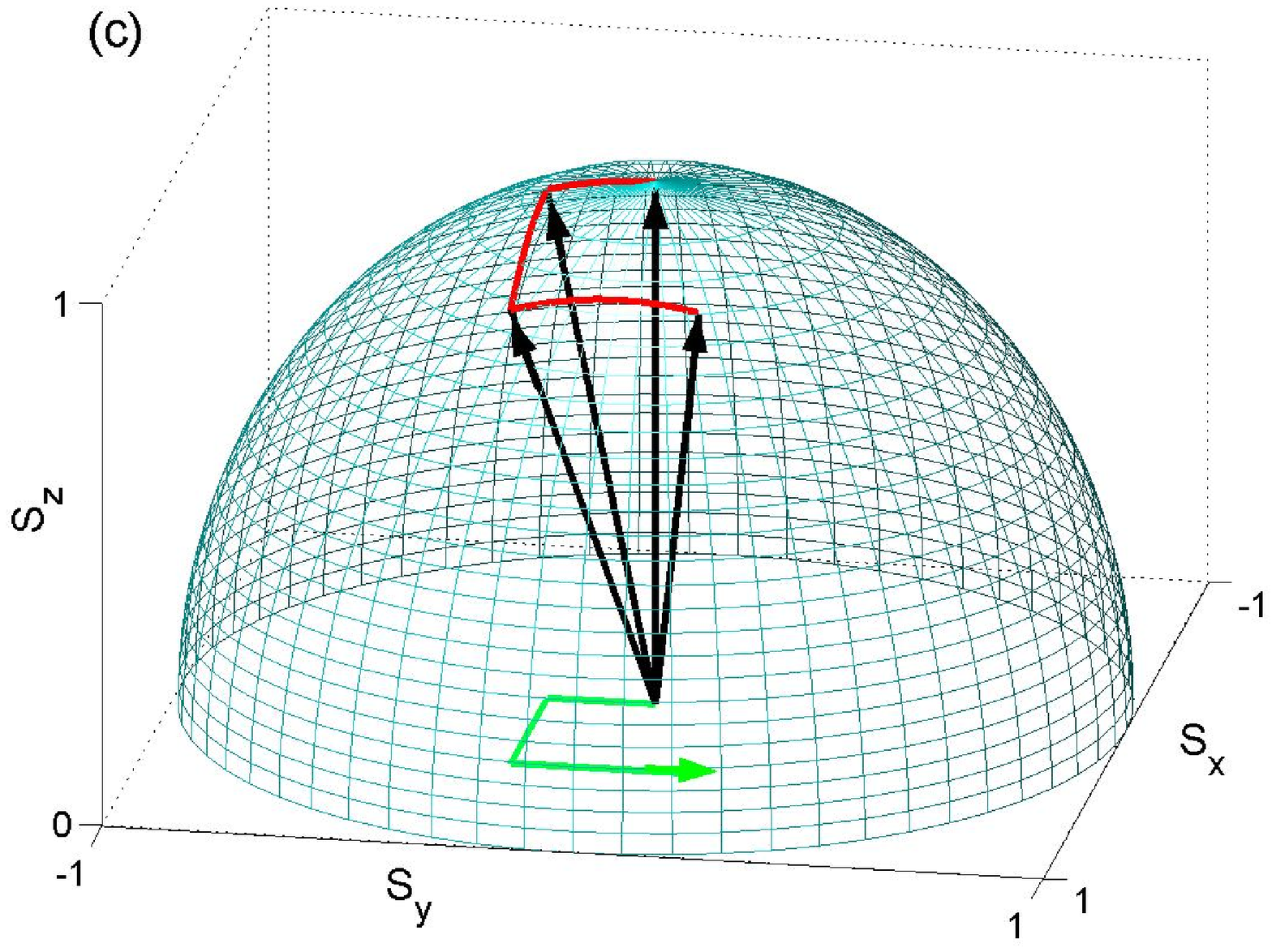}
\end{minipage}
\caption{(Color online) (a) Electron motion inside a quantum dot.
The trajectory consists of three straight segments $\gamma_1$,
$\gamma_2$, and  $\gamma_3$. (b) The corresponding spin evolution
on the $s_xs_y$ plane, which is projected from (c). (c) The spin
evolution induced by the Rashba spin-orbit interaction on the
3-dimensional unit sphere.}
\end{figure}

As an example of spin evolution induced by the Rashba
interaction, let us consider an electron confined inside a
quantum dot in Fig. 1(a), moving along the trajectory $\gamma$
which consists of three straight segments $\gamma_1$, $\gamma_2$,
and $\gamma_3$ with the respective lengths $l_1$, $l_2$, $l_3$
and the angles $w_1=\pi/2$, $w_2=\pi$, $w_3=3\pi/2$. The initial
spin state of this electron is polarized in $z$-direction, which
is represented by an arrow in Fig. 1(c). This arrow is projected
down to the origin $(0,0)$ on the $s_xs_y$ plane in Fig. 1(b).
When the electron starts its motion from the initial point $p$
along the segment $\gamma_1$ (Fig. 1(a)), its spin rotates around
the axis $\mathbf{N}_1=(1,0,0)$ and circumscribes an arc on the
3-dimensional sphere in Fig. 1(c). This curve is projected down
onto a straight line on the $s_xs_y$ plane. This line is parallel
to $\gamma_1$, but runs in a direction opposite to $\gamma_1$, as
shown in Fig. 1(b). After the first collision with the boundary
the electron further moves along the segment $\gamma_2$, while
its spin rotates around $\mathbf{N}_2=(0,1,0)$ and circumscribes
the second arc on the sphere in Fig. 1(c). The spin projection in
Fig. 1(b) now runs parallel to $\gamma_2$ in the direction
opposite to electron motion along $\gamma_2$. It is easy to see
that the spin evolution on other segments follows the same rule:
When an electron passes through the $j$-th segment in a certain
direction, the spin circumscribes on the 3D unit sphere an arc
around the axis $\mathbf{N}_j$. This arc, in its turn, is
projected onto the $s_xs_y$ plane as a straight line parallel to
the electron trajectory, but oppositely directed to it.

Further, let us proceed from the spin evolution on individual
trajectories to the spin evolution averaged over an ensemble of
trajectories. We consider an ensemble of electrons distributed
uniformly within a bounded area of a 2-dimensional
heterostructure. At $t=0$ these electrons have random outgoing
angles but the same spins polarized in $z$-direction. Let
$s_z^{(i)}(t)$ be the $z$ component of the electron spin at time
$t$ for the $i$-th trajectory. Then, in our numerical simulations
the integral in Eq. (\ref{eq20}) can be replaced by the sum,
\begin{equation}\label{eq27}
P^z_c(t)=\frac{1}{n}\sum_{i=1}^ns_z^{(i)}(t),
\end{equation}
where the sum runs over $n$ individual trajectories. The so
averaged spin polarization will be calculated in the following
five systems:
\begin{description}
\item[(a)] In 2-dimensional bulk (Fig. 2(a)) with the elastic
collision length $l$ distributed according to the Poisson law
Prob$(l)=e^{-l/l_m}/l_m $, where $l_m$ is the mean free path. It
is a stochastic open system. This is just the system where the
conventional D'yakonov-Perel' spin relaxation has to be observed.

\item[(b)] In a ballistic circular quantum dot of radius $1$ with
the smooth boundary in Fig. 2(b). Since the boundary is smooth,
the incident and reflection angles on the boundary are the same.
Since the system is ballistic, no scattering occurs inside the
dot. It is an integrable system with a high spatial symmetry.

\item[(c)] In a ballistic triangular quantum dot with
the smooth boundary  in Fig. 2(c). It is an integrable system of
lower symmetry  compared to the circular dot.

\item[(d)] In a generalized Sinai billiard with the
smooth boundary  in Fig. 2(d). It is a deterministic but strongly
chaotic system. The boundary geometry generates an ergodic
dynamics in the phase space.

\item[(e)] In a ballistic circular quantum dot like
Fig. 2(b), but with random reflections from the boundary. The
reflection angle takes random values between $-\pi/2$ and $\pi/2$
with respect to the boundary normal. It is a stochastic closed
system and corresponds to a quantum dot whose boundary is not
perfect in the scale of the electron Fermi wavelength.
\end{description}
The mean free path $l_m$ in bulk in Fig. 2(a) is set to $1$. The
sizes of the triangular and Sinai dots, as shown in Fig. 2(c) and
(d), are chosen to be $\sqrt{2\pi}\approx 2.5066$ and
$\sqrt{32\pi/(16-\pi)}\approx 2.7961$, such that these dots have
the same area $\pi$ as that of the circular dot in Fig. 2(b). We
will use the dimensionless time unit, such that during the time
interval 1 a particle moving with the Fermi velocity travels a
distance of the length 1.

\begin{figure}[htbp!]
\includegraphics[width=8.5cm]{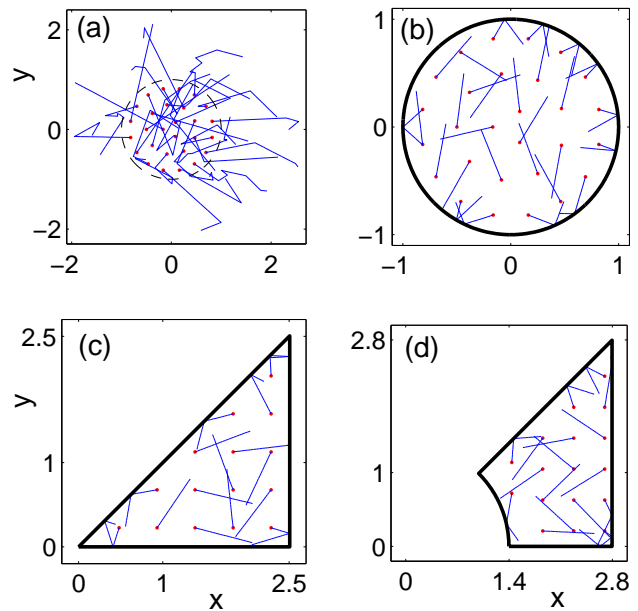}
\caption{(Color online) Electrons trajectories (solid lines) for
short time intervals: (a) in bulk, (b) circular quantum dot, (c)
triangular quantum dot, and (d) Sinai quantum dot.}
\end{figure}

\subsection{III.2 Results of the numerical simulations}

In Fig. 3 the time dependences of $P^z_c(t)$ for $2124$ electrons
in the open system (Fig. 2(a)) with $L_{so}=10$, $6$ and $2$ are
plotted by solid curves $C_1$, $C_2$, and $C_3$. One can see that
the relaxation time increases with $L_{so}$. These curves can be
fitted by the well known expression for the longitudinal DP
relaxation \cite{Ivchenko},
\begin{equation}\label{eq28}
P_{\tiny\mbox{DP}}(t)=\exp\left(\frac{-4\,t\,l_m}{L_{so}^2}\right),
\end{equation}
which is shown by the dashed curves in Fig. 3. This expression
was derived under the assumption of sufficiently large $L_{so}
\gg l_m$. For not so large $L_{so}$ the fitting is not good, as
it can be seen for the curve $C_3$ around its first drop at
$t=4$. In this regime the spin rotates rather fast, so that most
of the spins $s_z^{(i)}(t)$ evolve to negative values before the
electrons encounter their first collisions with impurities.
Therefore, $P^z_c(t)$ can evolve to a deep negative value within
a short time interval. But later on $P^z_c(t)$ approaches to the
asymptotic value $P^z_c=0$ (curve $C_3$ in Fig. 3). These results
from Monte Carlo simulations confirm the well known DP relaxation
in unbounded systems.

\begin{figure}[htbp!]
\begin{minipage}[b]{8.4cm}
\includegraphics[width=8.4cm]{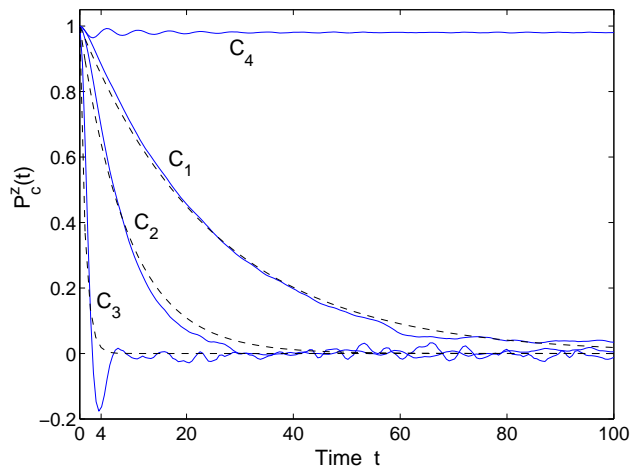}
\end{minipage}
\caption{Solid curves $C_1$, $C_2$, and $C_3$ represent the time
dependent polarization $P^z_c(t)$ for $2124$ particles in an
unbounded QW with $L_{so}=10$, $6$, $2$ and the mean free path
$l_m=1$. The particles were initially placed inside a circular
area of the radius $R=1$ and polarized in $z$-direction. The
dashed curves depict the DP relaxation calculated from Eq.
(\ref{eq28}). For comparison, curve $C_4$ shows $P^z_c(t)$ for
$2124$ particles confined inside a circular dot of the radius
$R=1$ and $L_{so}=10$.}
\end{figure}

If electrons are confined inside the smooth circular dot (Fig.
2(b)), the relaxation of $P^z_c(t)$ is considerably suppressed,
so that at large $L_{so}$ the spin polarization remains close to
$1$ at large times, as the curve $C_4$ in Fig. 3 demonstrates for
the case of $L_{so}=10$. At this regime, the suppression of
relaxation takes place in all other quantum dots, like the
circular dot with the rough boundary (curve $C_6$), the
triangular dot (curve $C_7$), and the Sinai billiard (curve
$C_8$) in Fig. 4. In all of these curves the $P^z_c(t)$ values
fall into the range between $0.97$ and $0.98$ at large times up
to $t=10^3$.

On the other hand, the spin polarization evolves very fast down
to $0$ if $L_{so}$ is smaller than the dot size. The
corresponding time dependence of $P^z_c(t)$ is similar to that
shown in Fig. 3 (curve C$_3$), with a sharp drop at the beginning
followed by oscillations around zero.

\begin{figure}[htbp!]
\begin{minipage}[b]{8.4cm}
\includegraphics[width=8.4cm]{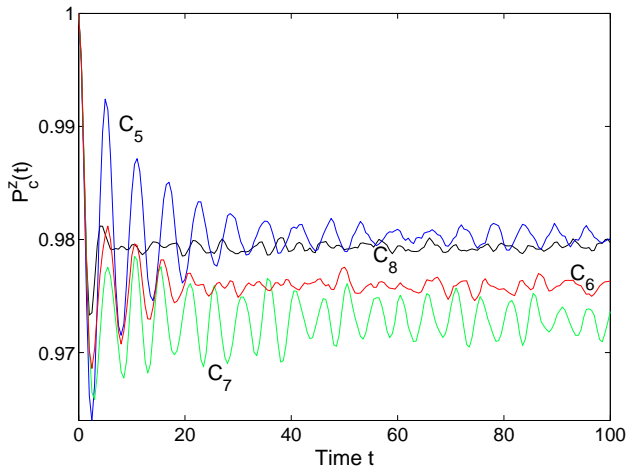}
\end{minipage}
\caption{(Color online) Time dependence of $P^z_c(t)$ for
$L_{so}=10$ in the smooth circular dot (curve $C_5$), the
circular dot with the rough boundary (curve $C_6$), the
triangular dot (curve $C_7$), and the Sinai billiard (curve
$C_8$).}
\end{figure}

\begin{figure}[htbp!]
\begin{minipage}[b]{8.4cm}
\includegraphics[width=8.4cm]{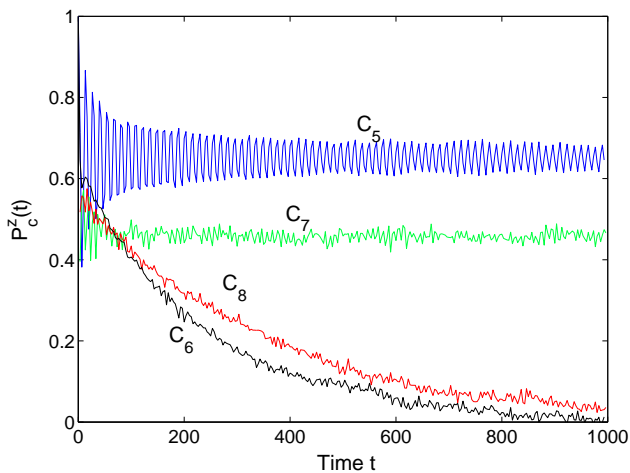}
\end{minipage}
\caption{(Color online) Time dependence of $P^z_c(t)$ for
$L_{so}=2$ in the smooth circular dot (curve $C_5$), the circular
dot with the rough boundary (curve $C_6$), the triangular dot
(curve $C_7$), and the Sinai billiard (curve $C_8$).}
\end{figure}

For an intermediate $L_{so}$ the spin relaxes according to
different scenarios, depending on the quantum dot geometry. As an
example, Fig. 5 shows the function $P^z_c(t)$ for various dot
geometries at $L_{so}=2$. After a fast initial drop, the
polarization further relaxes to 0 in the Sinai billiard (curve
$C_8$) and in the circular dot with the rough boundary (curve
$C_6$). However, in the smooth circular (curve $C_5$) and
triangular (curve $C_7$) dots this function oscillates around a
constant value at large times. It should be noted that in the
former two examples the spin polarization relaxes to zero at much
longer times than the DP relaxation time in the unbounded system
(Fig. 3), although the mean elastic scattering length there is
comparable to the dot size. The relaxation times for $C_6$ and
$C_8$ in Fig. 5 increase rapidly with higher $L_{so}$. Thus, at
$L_{so}=10$ we could not detect any systematic decrease of the
spin polarization in the Sinai billiard and rough circular dot,
up to $t=10^3$, which is by an order of magnitude larger than the
range plotted in Fig. 4.

An interesting feature of $P^z_c(t)$ in the regular systems, like
the triangular and smooth circular dots, is the apparent
oscillation of the polarization. It can be seen at Figs. 4 and 5,
although the oscillations in the latter figure are more profound
for the case of the circular dot, compared to almost vanishing
ripples in the triangle. These oscillations do not disappear at
large times and their amplitude increase with the strength of
SOI. We can not say much about their nature. Probably, they are
associated with the role of periodic trajectories in regular
systems. A special study is required to understand the origin and
characteristics of these oscillations.

\begin{figure}[htbp!]
\includegraphics[width=8cm]{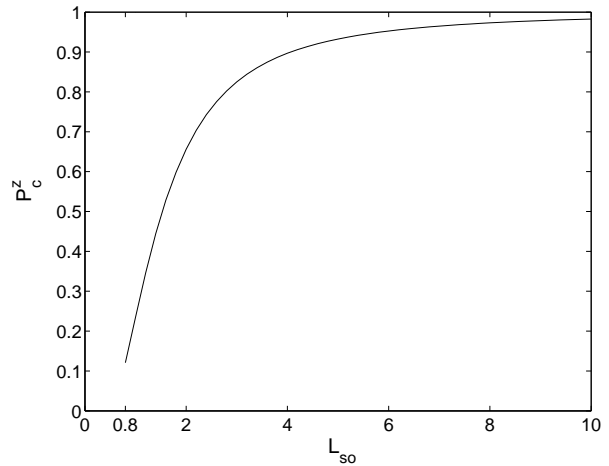}
\caption{The residual polarization $P^z_c$ vs the spin rotation
length $L_{so}$ for a smooth circular dot.}
\end{figure}

At long time the spin polarizations in both regular quantum dots
(triangle and smooth circle) in Fig. 4 and 5 oscillate around
certain nonzero residual values. These residual polarizations
$P^z_c$ are $L_{so}$ dependent, as plotted in Fig. 6 for the
circular dot.

\subsection{III.3 Spin evolution along individual trajectories}

The existence of the nonzero residual polarization in regular
quantum dots and long spin relaxation time in chaotic systems are
fundamentally distinct from the DP spin relaxation in the
boundless QW. Such a distinction is surprising, because at first
sight the spin walks on the sphere in Fig. 1(c) should be
randomized by scattering of particles from dot boundaries,
similar to randomization by impurity scattering in unbounded
systems. However, this simple point of view is wrong, because
there is an important difference between the impurity scattering
and the boundary scattering. For convenience, let us define the
scattering with a direction change smaller than $\pi/2$ as a
'forward' scattering and that larger than $\pi/2$ as a 'backward'
scattering. If the particles are isotropically scattered by an
impurity, half of them continue to move 'forward'. However, if
the particles are scattered by a smooth boundary, the particles
with incident angles between $-\pi/4$ to $\pi/4$ with respect to
the boundary normal will be reflected 'backward'. Since
statistically more particles hit the boundary within this range
of angles, the 'backward' scattering prevails in DQ's. This
property of particle scattering can also be extended to QD's with
rough boundaries. Further, according to Fig. 1, a 'backward'
particle motion is mapped onto a 'backward' spin walk. Hence, if
the spin moves away from the north pole in Fig. 1, after a
boundary scattering the spin is more likely bounced back towards
the north pole. Such a non-Markovian statistics of the spin walks
gives a clue for understanding the numerical results in
subsection III.2.

In order to make this argument more clear it is instructive to
study in detail the spin evolution along a single trajectory. As
described in Fig. 1, the spin motion on the unit sphere can be
projected onto the $s_xs_y$ plane. After a long time the spin
path on the sphere will cover a region and produce a certain
pattern on the $s_xs_y$ plane. In the circular dot this pattern
looks rather ordered. If the electron moves along a triangular
periodic trajectory (Fig. 7(a)), the pattern is a rounded
triangle (Fig. 7(i)). If the trajectories are hexahedral and
star-like (Fig. 7(b) and (c)), the corresponding patterns are a
rounded hexagon and a rounded star (Fig. 7(j) and (k)). If the
trajectories are non-periodic, e.g., Fig. 7(d), the pattern is a
disc (Fig. 7(l)). A common feature of these patterns is that they
have the same size, which is less than $1$ in the case of
$L_{so}=5$. These patterns are highly stable up to the
observation time $t=10^4$. It implies that the spin on the unit
sphere cannot move far away from the north pole, so that
$s_z^{(i)}(t)$ cannot take negative values. Our analysis of
various trajectories with various initial conditions has
confirmed this general feature of the spin evolution in the
circular dot. Hence, a non-zero $P^z_c$ in Fig. 6 at infinitely
long time is obviously expected.

In the triangular dot, two periodic and one non-periodic
trajectories are shown in Fig. 7(f), (g), and (h). The
corresponding spin patterns (Fig. 7(n), (o), and (p)) are less
symmetric and have less predictable sizes than those in the
circular dot. For the trajectory in Fig. 7(g) the pattern in Fig.
7(o) touches the circular border. Nevertheless, our investigation
shows that the patterns of most other trajectories are quite
stable up to the observation time 10$^4$ and do not touch the
boarder. On this reason the spin polarization being averaged over
trajectories is expected to relax to a positive residual value,
although this value is smaller than that in the smooth circular
dot.

\begin{figure}[htbp!]
\includegraphics[width=10cm]{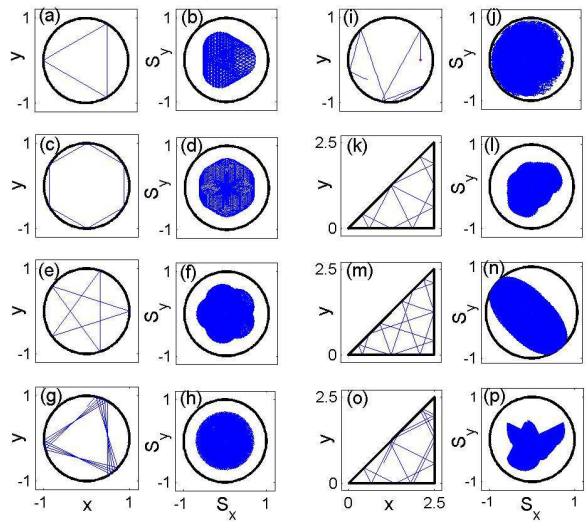}
\caption{Electron trajectories on the $xy$ plane ((a)-(h)) and
respective spin evolution patterns on the $s_xs_y$ plane
((i)-(p)) for $L_{so}=5$. (a), (b), and (c) Periodic triangular,
hexahedral, and star-like trajectories in the smooth circular
dot. (d) A non-periodic trajectory in the smooth circular dot.
(e) A stochastic trajectory in the rough circular dot. (f) and
(g) Two periodic trajectories in the triangular dot. (h) A
non-periodic trajectory in the triangular dot.
}
\end{figure}

In the circular dot with a rough boundary, the reflection angles
are stochastic, as shown in Fig. 7(e). Within the observation
time $t=10^3$ the corresponding spin pattern on the $s_xs_y$
plane has spread out to a much larger area (Fig. 7(m)) than those
in the smooth circular dot (Fig. 7(i)-(l)). Furthermore, the
pattern in (Fig. 7(m)) is still expanding. The corresponding spin
state on the 3D sphere can penetrate into the lower hemisphere
after $t=10^3$. However, it can return back to the north sphere
again. Therefore, the $z$ component of this spin state oscillates
between negative and positive values. When averaged over many
trajectories, such oscillations sum up to a relaxation curve,
similar to $C_6$ in Fig. 5.

In the Sinai billiard, the $s_xs_y$ pattern resembles that in the
rough circular dot. Consequently, the spin relaxation dynamics in
both cases have similar characteristics (curves $C_6$ and $C_8$
in Fig. 5).

A general trend seen from Fig. 7 is that the confinement of the
particle motion in QD's makes the spin to be also confined within
the upper hemisphere, if $L_{so}$ is larger than the size of the
QD's. For a smooth circular dot, this trend can be easily
understood from the 'backward' scattering effect described at the
beginning of this subsection. Since all trajectories in this case
have a simple geometry, one can easily see that particles are
more frequently scattered from the boundary in a 'backward'
direction. But although this argument holds for general bounded
systems, it is less evident for other QD's besides the smooth
circular dot. In general case, the trend toward the spin
confinement can be argued in a different way: As seen from Fig.
1, the projected spin path on the $s_xs_y$ plane in Fig. 1(b) is
more or less a rescaled curve of its particle trajectory in Fig.
1(a). But in reality the mapping from a trajectory to the
corresponding spin path is not simply a rescaling, because the
spin rotations on the sphere are non-commutative. For example, a
closed particle trajectory is in general mapped onto an open spin
path. However, if $L_{so}$ is large, the spin path is restricted
to a small part of the sphere. According to Eq.
(\ref{eq21}-\ref{eq22}), a closed particle trajectory produces an
open spin path of the linear size $\sim 1/L_{so}$, while the
distance between the initial and the end points of the path is
only $\sim 1/L_{so}^2$. The mapping between the trajectories and
the spin paths is then similar to a mapping between two Euclidean
spaces. Therefore, with the accuracy $1/L_{so}^2$, the spin paths
are the rescaled particle trajectories and those paths are
confined because the particle trajectories are confined. It
should be noted that such a tendency for the spin confinement
turns out to be strong even for not so large $L_{so}$, as one can
see from the spin dynamics shown in Fig. 5 for $L_{so}=2$.

The above argument about the spin confinement does not take into
account a long time evolution. Even at large $L_{so}$, small
corrections due to non-commutativity of spin walks will
accumulate in time. As a result, the spin can slowly drift toward
the lower hemisphere. The expanding pattern in Fig. 7(m) of the
rough circular dot is an example of such a long time behavior.
However, in contrast to that unstable pattern, the patterns from
regular systems (Fig. 7(i)-(p) besides (m)) remain stable in
time. This difference between the single trajectories of random
and regular systems is consistent with the spin relaxation curves
shown in Fig. 5.

Such a distinction between regular and chaotic systems follows
from fundamental properties of regular and chaotic systems. It
can be understood from consideration of periodic orbits. After a
particle runs along a periodic orbit $\gamma$ and completes a
period, its initial spin state $\chi$ will evolve to
$U_\gamma\chi$ with $U_\gamma=\exp[-i\Omega
\mathbf{R}\bm{\sigma}]$, which represents a rotation around the
axis $\mathbf{R}$ through the angle $2\Omega$. Both $\mathbf{R}$
and $\Omega$ are determined entirely by the geometry of $\gamma$
and by the value of $L_{so}$. After the particle repeats $w$
periods, all spin positions $(U_\gamma)^w\chi$, corresponding to
the end points of all periods $w=1,2,\cdots$, are located on a
closed circle. This circle can be obtained by rotating the north
pole around $\mathbf{R}$, if the initial $\chi$ is related to the
spin polarized in the north pole direction. The other points on
the periodic orbit are mapped onto spin states around this
circle. Taking many periodic orbits into account, one obtains a
set of different axes $\mathbf{R}$ and consequently a set of
circles passing through the north pole. Hence, when averaged over
all periodic orbits, spin spends more time in the upper
hemisphere. This means that at least the family of the periodic
orbits contributes to a nonzero residual polarization. How
significant is this contribution to the whole residual value
depends on the amount of the periodic orbits in a system, which
is quite different in regular and chaotic systems. In a regular
system the family of periodic orbits has a finitely positive
measure and a bundle of adjacent nearly periodic orbits. These
adjacent trajectories behave like periodic orbits if the time is
not too large, because their linear deviation in time from the
periodic orbits is small. On the contrary, the periodic orbits in
chaotic systems are of zero measure \cite{Gutzwiller}.
Furthermore, their adjacent trajectories deviate from them
exponentially fast. Therefore, with increasing time, the weight
of the periodic orbits and their adjacent trajectories becomes
exponentially small in chaotic systems, while it is a nonzero
value in regular systems. Hence, as long as we consider only
periodic orbits, the residual spin polarization has to be a
positive number for regular systems and zero for chaotic systems.

The individual trajectory study in a larger time scale carried
out in this subsection helps us to understand some of the results
in subsection III. 2. However, although the existence of the
nonzero residual polarization $P^z_c$ is apparent from Figs. 3-6,
one cannot exclude a possibility that $P^z_c$ will decay to zero
in a much larger time scale, since the numerical simulations in
all these figures are truncated within a finite time. Therefore,
we can not definitely answer the question whether the spin
polarization relaxes to zero at the infinitely long time, or to a
nonzero residual value. For the smooth circular dot the latter
alternative is corroborated by an analysis of the spin
polarization from the exact solution of the Schr\"{o}dinger
equation, as shown in the next section.

\section{IV Quantum mechanical polarization in the circular quantum dot}

Due to the time reversal symmetry, the quantized energy levels
$E_n$ of the Hamiltonian $H$ in Eq. (\ref{eq1}) are, at least, two
fold degenerate with the corresponding spinor eigenfunctions
$\varphi_{n\mathrm{a}}$, where $\mathrm{a}\in\{\pm\}$ is the
degeneracy index. In the basis of these states a normalized wave
function $\psi(\mathbf{r},t)$ can be expanded as
\begin{equation}\label{eq29}
\psi(\mathbf{r},t) =\sum_{n\mathrm{a}}
c_{n\mathrm{a}}\varphi_{n\mathrm{a}} (
\mathbf{r})\,e^{-i\,E_nt/\hbar}\,,
\end{equation}
with the coefficient
\begin{equation}\label{eq30}
c_{n\mathrm{a}}=\int
\varphi_{n\mathrm{a}}^{\dag}(\mathbf{r})\psi(\mathbf{r})\,d^{\,2}
r\,.
\end{equation}
The expression of $\psi(\mathbf{r},t)$ in Eq. (\ref{eq29})
differs from Eq. (\ref{eq3}) only by the degeneracy index
$\mathrm{a}$, which is explicitly written here for convenience of
our further analysis. Taking the notation
\begin{eqnarray}\label{eq31}
\psi_{n\mathrm{a}}(\mathbf{r},t)=c_{n\mathrm{a}}\varphi_{n\mathrm{a}}
(\mathbf{r})\,e^{-i\,E_nt/\hbar},
\end{eqnarray}
and
$\psi_{n\mathrm{a}}(\mathbf{r})=\psi_{n\mathrm{a}}(\mathbf{r},0)$,
the $z$ component of the quantum mechanical polarization in Eq.
(\ref{eq5}) can be expressed as
\begin{eqnarray}\label{eq32}
P^z(t)&=&\langle\psi(\mathbf{r},t)|\sigma^z|\psi(\mathbf{r},t)\rangle \nonumber \\
&=& \sum_{n\mathrm{a}\mathrm{b}}\int\psi_{n\mathrm{a}}^\dag(\mathbf{r})\sigma^z\psi_{n\mathrm{b}}(\mathbf{r})\, d^{\,2} r  \\
    &&+\!\!\!\!\!\sum_{n\neq m,\mathrm{a}\mathrm{b}}\!\int\psi_{n\mathrm{a}}^\dag(\mathbf{r})\sigma^z\psi_{m\mathrm{b}}(\mathbf{r}) \,
    e^{i(E_n-E_m)t/\hbar}\,d^{\,2} r  \nonumber.
\end{eqnarray}

The first sum in this equation is time independent, while the
second sum oscillates in time, so that its average over a
sufficiently long time interval turns to zero. It is interesting
to find out whether the former term coincides with the residual
polarization in Fig. 6. Such a coincidence is not evident because
the time dependent sum can give rise to large variations of
$P^z(t)$ after long time $t$. Moreover, the semiclassical theory
employed in the previous section can be not valid for times
larger than the mean distance between energy levels near the
Fermi energy. We can check such a coincidence at least for the
simple case of a circular dot with the smooth boundary, by
calculating the residual polarization
\begin{equation}\label{eq33}
P^z
=\sum_{n\mathrm{a}\mathrm{a'}}\int\psi_{n\mathrm{a}}^\dag(\mathbf{r})\sigma^z\psi_{n\mathrm{a'}}(\mathbf{r})\,
d^{\,2} r,
\end{equation}
because the analytic solution of the Schr\"{o}dinger equation
with the arbitrarily strong Rashba interaction is available
\cite{Tsitsishvili}. In this section only the key steps of the
calculation are presented, while the calculation in detail is
shown in the Appendix.

Let us consider a circular quantum dot of radius $R$ with the
confining potential
\begin{equation}\label{eq34}
V(\rho)=\left\{\begin{array}{ll}
0 & \mbox{ for }0 \leq \rho\leq R \\ \infty & \mbox{ for }R >\rho
\end{array} \right. ,
\end{equation}
written as a function of the polar coordinates
$\mathbf{r}=\mathbf{r}(\rho,\phi)$. The eigenfunctions of the
n-th eigenvalue $E_n$ are \cite{Tsitsishvili}
\begin{equation}\label{eq35}
\varphi_{n+}(\mathbf{r})=\left(\begin{array}{c} e^{i\nu\phi}f_\nu(\rho) \\
e^{i(\nu+1)\phi}g_{\nu+1}(\rho)
\end{array} \right)
\end{equation}
and
\begin{equation}\label{eq36}
\varphi_{n-}(\mathbf{r})=\left(\begin{array}{c} e^{-i(\nu+1)\phi}g^*_{\nu+1}(\rho) \\
-e^{-i\nu\phi}f^*_\nu(\rho)
\end{array} \right),
\end{equation}
where the function
\begin{equation}\label{eq37}
\left(\begin{array}{c}f_\nu(\xi) \\
g_\nu(\xi)\end{array}\right)
=d\left(\begin{array}{c} -a_\nu J_\nu(k_+\xi)+J_\nu(k_-\xi) \\
a_{\nu-1} J_\nu(k_+\xi)+J_\nu(k_-\xi)\end{array} \right),
\end{equation}
contains the $\nu$-th order Bessel functions of the first kind
$J_\nu(\xi)$, the normalization constant $d$, the parameters
\begin{equation}\label{eq38}
a_\nu=\frac{J_\nu(k_-)}{J_\nu(k_+)}=-\frac{J_{\nu+1}(k_-)}{J_{\nu+1}(k_+)},
\end{equation}
the wave numbers
\begin{equation}\label{eq39}
k_\pm=\frac{\sqrt{b^{\,2}+4\varepsilon}\mp b}{2},
\end{equation}
and the index
\begin{equation}\label{eq40}
\nu=\lambda-3/2\hspace{0.3cm}\mbox{ with
}\hspace{0.3cm}\lambda=1,\,2,\,\cdots.
\end{equation}
Therein, the dimensionless parameters $\xi=\rho/R$,
$\varepsilon=2m^*ER^2/\hbar^2$, and $b=2\alpha_R m^*R/\hbar^2$
have been used. The wave numbers $k_\pm$ are quantized because
the energy levels $\varepsilon$ are determined by the zeros of
the function
\begin{equation}\label{eq41}
Z_\nu(\varepsilon):=J_{\nu}(k_-)J_{\nu+1}(k_+)+J_{\nu}(k_+)J_{\nu+1}(k_-).
\end{equation}

We chose the plane wave
\begin{equation}\label{eq42}
\psi(\mathbf{r})=\left(\begin{array}{c} 1
\\ 0\end{array} \right)e^{i\mathbf{k} \mathbf{r}}
\end{equation}
as the initial state. After inserting $\varphi_{n+}(\mathbf{r})$
from Eq. (\ref{eq35}) and $\varphi_{n-}(\mathbf{r})$ from Eq.
(\ref{eq36}) together with Eq. (\ref{eq31}) into Eq. (\ref{eq33})
and averaging over directions of the vector $\mathbf{k}$ we
obtain
\begin{equation}\label{eq43}
P^z=2\pi\sum_n \left(|c_{n+}|^2-|c_{n-}|^2\right)(F_n-G_n),
\end{equation}
with
\begin{eqnarray}\label{eq44}
&&F_n=d^{\,2}\left[\,a_\nu^2\,I_\nu^{(1)}-2a_\nu\,I_\nu^{(2)}+I_\nu^{(3)}\right] \nonumber\\
&&G_n=d^{\,2}\left[\,a_\nu^2\,I_{\nu+1}^{(1)}+2a_{\nu}\,I_{\nu+1}^{(2)}+I_{\nu+1}^{(3)}\right],
\end{eqnarray}
where the coefficients $I_\nu^{(1)}$, $I_\nu^{(2)}$, and
$I_\nu^{(3)}$ are presented in Eq. (\ref{eq63}). The coefficients
$|c_{n\pm}|^2$ in Eq. (\ref{eq43}) can be written as
\begin{eqnarray}\label{eq45}
&&|c_{n+}|^2=4\pi^2\,d^{\,2}\,\left(\,-a_\nu\,I_\nu^{(4)}+I_\nu^{(5)}\,\right)^2 \nonumber\\
&&|c_{n-}|^2=4\pi^2\,d^{\,2}\,\left(\;a_\nu\,I_{\nu+1}^{(4)}+I_{\nu+1}^{(5)}\,\right)^2,
\end{eqnarray}
with the coefficients $I_\nu^{(4)}$ and $I_\nu^{(5)}$ given by
Eq. (\ref{eq66}). Using the dimensionless units, one has the
radius $R=1$, the coupling constant $b=2/L_{so}$, and the wave
number $k =2\pi R/\lambda$, where $\lambda$ is the electron
wavelength. Hence the semiclassical range of parameters
corresponds to $k \gg 1$.

The residual polarization calculated from Eq. (\ref{eq43}) is
shown in Fig. 8. The $P^z$ curves for $k=20$, $30$, and $40$ are
very close to each other and merge into the dashed curve. This
curve coincides with the residual polarization obtained from the
semiclassical simulations in the previous section (Fig. 6). For
$k=5$, $1$, and $0.1$, the curves are plotted in the dotted,
solid, and dash-dotted curves, respectively. All the curves, as
expected, have the common asymptotic value $1$ in the case of the
weak spin-orbit coupling $L_{so} \rightarrow \infty$. In the
opposite limit, $L_{so} \rightarrow 0$, the behavior of $P^z$ is
nonanalytic and not much revealing. The strong oscillations in
this limit increase with smaller wave numbers and signal about
the appearance of large quantum beats in $P^z (t)$. This regime
of $L_{so}$ is not interesting from the practical point of view
because it implies unphysically large values of $\alpha_R$ for
the typical dot radius $R=500 nm$. In the practically important
regime of $L_{so} \geq 1$ we note an apparent dependence of $P^z$
on $k$ at $k\leq 5$. This is a quantum effect which is not
observed in our semiclassical simulations. In semiclassics the
particle velocity determines the speed with which $P^z (t)$
approaches to the residual value $P^z$, but not this value
itself.

\begin{figure}[htbp!]
\includegraphics[width=8cm]{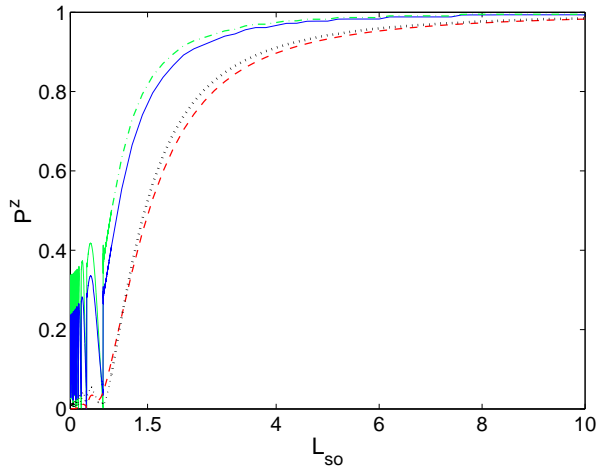}
\caption{(Color online) The residual spin polarization $P^z$ vs
$L_{so}$ with $k=20$, $30$, $40$, (dashed), $k=5$ (dotted), $k=1$
(solid), and $k=0.1$ (dash-dotted). The dashed curve coincides
with the curve from Fig. 6.}
\end{figure}

\section{V Discussion}

Summarizing the above results of the semiclassical Monte Carlo
simulations and quantum mechanical calculations we can draw the
following picture of the spin evolution in semiclassical quantum
dots. In the dots with regular classical dynamics the spin
polarization does not decay to zero at long time and its residual
value coincides with the quantum mechanical spin polarization
averaged over an infinitely long time interval. At least, we were
able to check such a coincidence for the circular dot. On the
other hand, in dots with chaotic or random dynamics the spin
polarization relaxes to zero with the relaxation time much larger
than the DP relaxation time in unbounded quantum wells. Such a
decay down to zero can not be understood from the general quantum
mechanical expression in Eq. (\ref{eq32}), because it implies
that the average of $P^z (t)$ over an asymptotically long time
interval is zero. However, Eq. (\ref{eq32}) predicts that this
average is given by the first term in Eq. (\ref{eq32}), which is
nonzero in general. Obviously, this contradiction is associated
with quantum mechanical effects, which indicates that the
semiclassical approximation is insufficient for analysis of the
long time polarization evolution. In disordered mesoscopic
systems the statistics of their energy spectrum together with
quantum interference effects give rise to the so called quantum
dynamical echo \cite{Prigodin} which can contribute to the spin
evolution at large times. This problem needs further study.

The predicted spin evolution can be measured experimentally. For
an InAs dot doped up to 10$^{11}$cm$^{-2}$, the time unit in
Figs. 3-5 is about 1 ps if the dot size is $L=0.5 \mu m$. Hence,
the spin polarization saturates to its residual value during
first 20 ps and for $L_{so}=1 \mu m$ the difference in the long
time evolution between chaotic and regular dots can be observed
in the nanosecond range. In order to suppress all inelastic spin
relaxation mechanisms \cite{Khaetskii, Woods, Merkulov, Semenov},
the measurement must be done at sufficiently low temperatures.
The Rashba spin-orbit interaction can be strong in InAs based
heterostructures, with $L_{so}$ down to several hundreds nm.
Moreover, it can be tuned in a wide interval by varying the gate
voltage \cite{Nitta}.

In conclusion, we performed path integral semiclassical
simulations of spin evolution controlled by the Rashba spin-orbit
interaction in quantum dots of various shapes. Our calculations
revealed that the spin polarization dynamics in QD's is quite
different from the D'yakonov-Perel' spin relaxation in bulk 2D
systems. Such a distinction is not expected from the simple
picture of the spin random walk, in particular when the rate of
electron elastic scattering on impurities in bulk is equal to the
mean frequency of electron scattering from the dot boundaries. We
have also found an important distinction between long time spin
evolutions in classically chaotic and regular systems. In the
former case the spin polarization relaxes to zero within
relaxation time much larger than the DP relaxation, while in the
latter case it evolves to a time independent residual value. This
value decreases with the stronger spin orbit interaction. We also
analyzed the general quantum mechanical expression for the time
dependent spin polarization. Using the exact solutions of the
Schr\"{o}dinger equation with Rashba SOI for a circular dot, we
calculated the average of the spin polarization over an
infinitely long time interval and compared the result with the
residual polarization from the Monte Carlo simulations. We found
that the residual values from these two approaches coincide,
which confirms the results from the semiclassical simulations. On
this basis, we conjecture that the nonzero residual value is a
general property of regular systems. On the other hand, the spin
relaxation down to zero in the Sinai billiard and circular dot
with the rough boundary contradicts to what have to be expected
from quantum mechanics. The long time memory due to the
mesoscopic spin echo is assumed to be responsible for this
contradiction.

\section{Appendix}
This appendix demonstrates a quantum mechanical calculation of
the residual polarization $P^z$, as it is defined in Eq.
(\ref{eq33}). The calculation of the exact eigenfunctions of the
Hamiltonian in Eq. (\ref{eq1}) for the circular quantum dot can
be found in Ref. \cite{Tsitsishvili}, which is summarized in the
following Eqs. (\ref{eq46}-\ref{eq50}).

In order to calculate the residual polarization (\ref{eq33}), the
wave function $\psi_{na}(\mathbf{r})$ is expanded in the basis of
the eigenfunctions given by Eqs. (\ref{eq35}-\ref{eq36}). We note
that for a symmetric presentation, the functions $f_\nu$ and
$g_\nu$ have different definitions from those in Ref.
\cite{Tsitsishvili}. Inserting Eqs. (\ref{eq35}-\ref{eq36}) into
the corresponding Schr\"{o}dinger equation we obtain the equation
for $f_\nu$ and $g_\nu$ in terms of the dimensionless parameters
$\xi, \varepsilon$, and $b$ defined in the previous section,
\begin{eqnarray}\label{eq46}
&&[\triangle_\nu +\varepsilon]\,f_\nu(\xi)-b\, \nabla_{-(\nu+1)} \,g_{\nu+1}(\xi) =0, \nonumber \\
&&[\triangle_{\nu+1}
+\varepsilon]\,g_{\nu+1}(\xi)-b\,\nabla_{+\nu}\,f_\nu(\xi) =0,
\end{eqnarray}
with the Laplacian
\begin{equation}\label{eq47}
\triangle_\nu=\frac{1}{\xi}\frac{d}{d\xi}\left(\xi\frac{d}{d\xi}\right)-\frac{\nu^2}{\xi^2}
\end{equation}
and the nabla operator
\begin{equation}\label{eq48}
\nabla_{\pm\nu}=\pm\left(\frac{d}{d\xi}\right)-\frac{\nu}{\xi}.
\end{equation}
The solutions $(f_\nu(\xi),g_\nu(\xi))$ of these equations are
\begin{equation}\label{eq49}
\left(\begin{array}{c}f_\nu(\xi) \\
g_\nu(\xi)\end{array}\right)
=d\left(\begin{array}{c} -a_\nu J_\nu(k_+\xi)+J_\nu(k_-\xi) \\
a_{\nu-1} J_\nu(k_+\xi)+J_\nu(k_-\xi)\end{array} \right),
\end{equation}
with the normalization constant $d$, the factors $a_\nu$ given by
Eq. (\ref{eq38}), and the wave vectors $k_\pm$ from Eq.
(\ref{eq39}). These wave vectors obey the relations
\begin{eqnarray}\label{eq50}
&&k_+k_-=\varepsilon,\hspace{0.3cm} k_+-k_-=-b, \nonumber \\
&&\mbox{and } k_++k_-=\sqrt{b^2+4\varepsilon}.
\end{eqnarray}

The quantized dimensionless energies $\varepsilon$ are determined
by the zeros of the function in Eq. (\ref{eq41}). This function
stems from the determinant of the equation system in Eq.
(\ref{eq46}) with the boundary conditions
$f_\nu(\xi)=g_\nu(\xi)=0$ at $\xi=1$. Given a coupling constant
$b$, the $n$-th quantized value $\varepsilon_n$ with
$n=n(\lambda,\mu)$ is determined by the $\mu$-th zero of
$Z_\nu(\varepsilon)$, where $\nu$ and $\lambda$ are related by
Eq. (\ref{eq40}). The allowed wave numbers $k_\pm$ are given by
Eq. (\ref{eq39}) with $\varepsilon=\varepsilon_n$. They
correspond to the two degenerate eigenstates of the n-th energy
level. The first root of the function $Z_\nu(\varepsilon)$ is
zero for $\nu=1/2,3/2,...$ and is a positive value for $\nu=-1/2$
(see Fig. 9). The larger the value of $b $, the larger the second
root of $Z_\nu(\varepsilon)$.

\begin{figure}[htbp!]
\includegraphics[width=8.5cm]{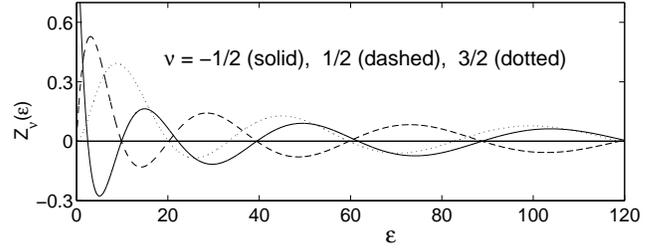}
\caption{The function $Z_\nu(\varepsilon)$ for $\nu=-1/2$, $1/2$,
and $3/2$. This function is singular at $\varepsilon=0$ for $\nu=-1/2$.}
\end{figure}
Substituting the wave functions in Eqs. (\ref{eq35}-\ref{eq36})
into Eq. (\ref{eq33}) we obtain the residual polarization in the
form
\begin{eqnarray}\label{eq51}
P^z&=&\sum_n\int_0^1\int_0^{2\pi}
\left[\left(|c_{n+}|^2-|c_{n-}|^2\right)
\left(f_{\nu}^2(\xi)-g_{\nu+1}^2(\xi)\right)\right. \nonumber\\
&&+2c_{n+}^*c_{n-}f_{\nu}(\xi)g_{\nu+1}(\xi)e^{-i(2\nu+1)\phi} \nonumber\\
&&\left .
+2c_{n-}^*c_{n+}f_{\nu}(\xi)g_{\nu+1}(\xi)e^{i(2\nu+1)\phi
}\right]\,d\phi\,\xi\,d\xi.
\end{eqnarray}

For the initial wave function given by Eq. (\ref{eq42}), the
constants $c_{n+}$ can be expressed as
\begin{eqnarray}\label{eq52}
c_{n+}&=&\int \left(\begin{array}{c} \varphi_{n+}^{(1)}(\mathbf{r}) \\
\varphi_{n+}^{(2)}(\mathbf{r})\end{array}\right)^\dag\,\left(\begin{array}{c}
1 \\ 0\end{array}
\right)e^{i\mathbf{k} \mathbf{r}}\,d^{\,2} r \nonumber \\
&=&\int \varphi_{n+}^{(1)\,*}(\mathbf{r})\,e^{i\mathbf{k} \mathbf{r}}\,d^{\,2} r \nonumber\\
&=& \int_0^{2\pi} \int_0^1
e^{i\left[k\xi\cos(\phi-\theta)-\nu\phi\right]}f_{\nu}(\xi)\xi\,d\xi\,d\phi,
\end{eqnarray}
where $\phi$ and $\theta$ stand for the angles of the vectors
$\mathbf{r}$ and $\mathbf{k}$ with respect to the positive x-axis
and $k=|\mathbf{k}|$. After the shift of the angular variable
from $\phi-\theta$ to $\phi$ the above integral transforms to
\begin{equation}\label{eq53}
c_{n+}=e^{-i\nu\theta}\int_0^{2\pi}\int_0^1
e^{i\left[k\xi\cos(\phi)-\nu\phi\right]}
f_{\nu}(\xi)\xi\,d\xi\,d\phi .
\end{equation}
Substituting $t=\phi+\pi/2$ and $m=\nu$ into the integral
representation of the Bessel function \cite{Gradshteyn},
\begin{eqnarray}\label{eq54}
J_m(z)=\frac{1}{2\pi}\int_{-\pi}^{\pi}e^{i\,[z\sin(t)-mt]}\,dt,
\end{eqnarray}
we obtain
\begin{eqnarray}\label{eq55}
2\pi\,e^{i\nu\pi/2}J_{\nu}(z)=\int_0^{2\pi}e^{i\,[z\cos(\phi)-\nu\phi]d\phi}.
\end{eqnarray}
By using this identity, Eq. (\ref{eq53}) can be written as
\begin{eqnarray}\label{eq56}
c_{n+}&=&2\pi\, e^{i\nu(\pi/2-\theta)}\int_0^1
J_{\nu}(k\xi)\,f_{\nu}(\xi)\xi\,d\xi.
\end{eqnarray}
By analogy, one has
\begin{eqnarray}\label{eq57}
c_{n-}=2\pi\, e^{i(\nu+1)(\pi/2-\theta)}\int_0^1
J_{\nu+1}(k\xi)\,g_{\nu+1}(\xi)\xi\,d\xi.
\end{eqnarray}

After integrating Eq. (\ref{eq51}) over $\phi$ and the direction
$\theta$ of $\mathbf{k}$ (integration over $\theta$ is similar to
that over $\theta_\mathbf{k}$ in Eq. (\ref{eq20})), the second
and third terms in Eq. (\ref{eq51}) vanish and only the first
term remains. Introducing the parameters
\begin{equation}\label{eq58}
F_n=\int_0^1 f_{\nu}^2(\xi)\xi \,d\xi \hspace{0.3cm}\mbox{ and }
\hspace{0.3cm}G_n=\int_0^1 g_{\nu+1}^2(\xi)\xi \,d\xi,
\end{equation}
the final expression for the residual polarization can be written
as
\begin{equation}\label{eq59}
P^z=\frac{\sum_n \left(|c_{n+}|^2-|c_{n-}|^2\right)(F_n-G_n)}
{\sum_n \left(|c_{n+}|^2+|c_{n-}|^2\right)(F_n+G_n)}.
\end{equation}
For numerical calculations we explicitly wrote the norm of the
normalized wave function $\psi(\mathbf{r},t)$ in the denominator.
In this form the expression in Eq. (\ref{eq59}) is also valid for
non-normalized wave functions, because the normalization
constants $d$ in the numerator and denominator are canceled with
each other.

The polarization $P^z$ in Eq. (\ref{eq59}) is determined by the
four integrals $c_{n\pm}$, $F_n$, and $G_n$. They can be
calculated by using the formula \cite{Lebedev}
\begin{eqnarray}
&&\int_0^{\,l} \xi\, J_\nu(\lambda \,\xi)J_\nu(\kappa \,\xi)\,d\xi \label{eq60} \\
&&\;\;\;\;=\frac{l\, [\kappa J_\nu(\lambda \,l)J_{\nu+1}(\kappa
\,l)-\lambda J_\nu(\kappa\,l) J_{\nu+1}(\lambda
\,l)]}{\kappa^2-\lambda^2}. \nonumber
\end{eqnarray}
Consequently, the integrals in Eq. (\ref{eq58}) can be written in
the closed form
\begin{eqnarray}
&&F_n=d^{\,2}\left[\,a_\nu^2\,I_\nu^{(1)}-2a_\nu\,I_\nu^{(2)}+I_\nu^{(3)}\right], \label{eq61} \\
&&G_n=d^{\,2}\left[\,a_\nu^2\,I_{\nu+1}^{(1)}+2a_{\nu}\,I_{\nu+1}^{(2)}+I_{\nu+1}^{(3)}\right],\label{eq62}
\end{eqnarray}
with
\begin{eqnarray}
&&I_\nu^{(1)}=\frac{J_\nu(k_+)^2+J_{\nu+1}(k_+)^2}{2}-\frac{\nu
J_\nu(k_+)J_{\nu+1}(k_+)}{k_+}, \nonumber\\
&&I_\nu^{(2)}=\frac{k_-J_\nu(k_+)J_{\nu+1}(k_-)-k_+J_\nu(k_-)J_{\nu+1}(k_+)}{k_-^2-k_+^2}, \nonumber\\
&& I_\nu^{(3)}=\frac{J_\nu(k_-)^2+J_{\nu+1}(k_-)^2}{2}-\frac{\nu
J_\nu(k_-)J_{\nu+1}(k_-)}{k_-}. \nonumber\\ \label{eq63}
\end{eqnarray}

By analogy, calculating the integrals in Eqs.
(\ref{eq56}-\ref{eq57}) we obtain
\begin{eqnarray}
&&|c_{n+}|^2=4\pi^2\,d^{\,2}\,\left(\,-a_\nu\,I_\nu^{(4)}+I_\nu^{(5)}\,\right)^2, \label{eq64}\\
&&|c_{n-}|^2=4\pi^2\,d^{\,2}\,\left(\;a_\nu\,I_{\nu+1}^{(4)}+I_{\nu+1}^{(5)}\,\right)^2,\label{eq65}
\end{eqnarray}
with
\begin{eqnarray}
I_\nu^{(4)}&=&\frac{k_+J_\nu(k)J_{\nu+1}(k_+)-kJ_\nu(k_+)J_{\nu+1}(k)}{k_+^2-k^2}, \nonumber\\
I_\nu^{(5)}&=&\frac{k_-J_\nu(k)J_{\nu+1}(k_-)-kJ_\nu(k_-)J_{\nu+1}(k)}{k_-^2-k^2}. \nonumber\\
\label{eq66}
\end{eqnarray}

For small $b$ the spin polarization approaches to $P^z=1$, as it
must be in the absence of the spin-orbit interaction. It follows
from the relation $k_--k_+=b \approx 0$ in Eq. (\ref{eq50}),
which results in $f_\nu(\xi)\approx 0$, according to the
definition in Eq. (\ref{eq49}). Hence, the two quantities
$|c_{n+}|$ and $F_n$, which contain $f_\nu(\xi)$, vanish in
$P^z$. Therefore, the sums in the numerator and denominator of
$P^z$ become the same, which gives rise to $P^z=1$.

For large $b$, we have $P^z\rightarrow 0$, which is due to the
large difference between $k_+$ and $k_-$, namely, $k_--k_+=b \gg
1$. According to the asymptotic behavior \cite{Lebedev}
\begin{equation}\label{eq67}
J_\nu(x)=\sqrt{\frac{2}{\pi
x}}\cos\left(x-\frac{\pi}{2}\left(\nu+\frac{1}{2}\right)\right)
+O\left(\frac{1}{x}\right)
\end{equation}
of the Bessel function at large $x$, the magnitude of the
oscillating function $J_\nu(k_+)$ is much larger than
$J_\nu(k_-)$ by the order of $\sqrt{k_-/k_+}$. Therefore, the
leading terms of $I_\nu^{(4)}$ and $I_\nu^{(5)}$ in Eq.
(\ref{eq66}) behave like
\begin{eqnarray}
I_\nu^{(4)}&\sim& \frac{J_\nu(k_+)J_{\nu+1}(k)}{k},  \nonumber \\
I_\nu^{(5)}&\sim&\frac{J_\nu(k)J_{\nu+1}(k_-)}{k_-}. \label{eq68}
\end{eqnarray}
The first term is much larger than the second one. Consequently,
both $c_{n+}$ and $c_{n-}$ are dominated by $I_\nu^{(4)}$ and
have the same limit for large $b$. By analogy, $F_n$ and $G_n$
also have the same limit. Therefore, both $|c_{n+}|^2-|c_{n-}|^2$
and $F_n-G_n$ in the numerator of Eq. (\ref{eq59}) become small
and hence $P^z=0$ for $b\rightarrow \infty$.

Acknowledgement: This work was supported by RFBR No. 03-02-17452
and the Swedish Royal Academy of Science; A.G.M. acknowledges the
hospitality of NCTS in Taiwan. C.-H.C. acknowledges the
hospitality of Lund university in Sweden and the Science
Institute in the university of Iceland.

\end{document}